\documentclass[10pt,conference]{IEEEtran}

\usepackage{cite}
\usepackage{amsmath,amssymb}
\usepackage{graphicx}
\usepackage{booktabs}
\usepackage{xcolor}
\usepackage{svg}
\usepackage{url}
\usepackage{comment}               
\usepackage[hidelinks]{hyperref}   
\usepackage{orcidlink}
\begin{document}

\title{RiskyPy: Toward Model-Calibrated Prompt Generation for Joint Security and Functionality Evaluation of LLM-Generated Code}




\author{
\IEEEauthorblockN{
Zacharie Chenail-Larcher\textsuperscript{1,*}\,\orcidlink{0009-0002-7464-6659}\quad
Daniele Lunghi\textsuperscript{2,*}\,\orcidlink{0000-0003-1324-981X}\\[0.5ex]
Melissa Tessa\textsuperscript{2}\,\orcidlink{0009-0008-7525-5522}\quad
Quentin Stiévenart\textsuperscript{3}\,\orcidlink{0000-0001-9985-9808}\\[0.5ex]
Jacques Klein\textsuperscript{2}\,\orcidlink{0000-0003-4052-475X}\quad
Tegawendé F. Bissyandé\textsuperscript{2}\,\orcidlink{0000-0001-7270-9869}\\[0.5ex]
Florent Avellaneda\textsuperscript{3}\,\orcidlink{0000-0003-1030-5388}\quad
Naouel Moha\textsuperscript{1}\,\orcidlink{0000-0001-9252-9937}
}
\IEEEauthorblockA{
\textsuperscript{1}École de technologie supérieure, Montréal, Québec, Canada\\
\textsuperscript{2}University of Luxembourg, Luxembourg, Luxembourg\\
\textsuperscript{3}Université du Québec à Montréal, Montréal, Québec, Canada\\
\textsuperscript{*}The first two authors contributed equally to this work and share first authorship.
}
}

\maketitle

\begin{abstract}
Evaluating approaches that improve the security of LLM code generation requires datasets that jointly assess security and functionality and remain adapted to the capabilities of the models being studied. However, existing benchmarks are costly to produce and are generally not calibrated to a chosen range of model capabilities. We introduce RiskyPy, an automated pipeline that generates security-oriented code-generation datasets calibrated to model proficiency using functionality and vulnerability canaries. We instantiate RiskyPy for Python using domains covering 18 CWEs from the 2025 CWE Top 25, producing 101 prompts. Experiments on Qwen3-4B and Seed-Coder-8B show that the generated tasks remain functionally accessible, with 57–72\% of prompts yielding functional code, while only 5–10\% of generations are free of Bandit findings, leaving room for joint evaluation of security and functionality of security improvement approaches.
\end{abstract}


\section{Introduction}
 Large Language Models (LLMs) show strong aptitude in code generation \cite{chen2021codex, hou2024llm4se, jimenez2024swebench}. Increasingly, they can handle complex tasks and produce functional code, i.e., code that completes the required task without errors.
However, their implementations are not always secure and may include vulnerabilities \cite{pearce2022asleep, perry2023insecure, fu2025copilot}. To address this issue, it is essential to develop approaches to protect against vulnerabilities.

Evaluating these approaches requires benchmarks that jointly assess the security and functionality of generated code. Otherwise, an approach may improve security by breaking the implementation \cite{dai2026rethinking, tessa2026secure}. Tasks where LLMs produce functional but insecure code are particularly useful, as they allow us to verify that approaches improve security while preserving functionality. Moreover, since such approaches are often evaluated across models of different strengths \cite{he2024safecoder, fu2024codeguard}, datasets should remain functionally accessible to weaker models while still exposing insecure implementations from stronger ones.

Creating such datasets remains difficult. Fixed benchmarks may eventually saturate \cite{ott2022saturation} and producing new datasets is costly. Writing tasks manually is time-consuming \cite{peng2025cweval, vero2025baxbench}, expanding expert-authored coding tasks still requires manual work \cite{nie2025secodeplt}, and mining repositories is limited by available vulnerabilities and risks data leakage \cite{zhao2025susvibes, lee2025secbench}. Furthermore, several datasets are derived from the same underlying coding tasks, potentially limiting their diversity~\cite{tony2023llmseceval, wang2026codeseceval}.

Existing dataset construction methods also rarely calibrate tasks to model capabilities, and use only frontier models to identify prompts (coding tasks expressed in natural language) that lead to vulnerable code \cite{hajipour2024codelmsec} or validate the tasks \cite{vonarx2026autobaxbuilder}.
To the best of our knowledge, none specifically selects tasks that weaker models can implement functionally while stronger models still tend toward insecure implementations. This is particularly relevant for research, as white-box approaches require open weights and larger models may be computationally inaccessible. For example, Qwen2.5 7B produces functional code for only about 5\% of tasks in the subset of AutoBaxBench marked by the authors as Easy \cite{vonarx2026autobaxbuilder}.
This highlights the need for an automated approach to generate security-oriented code generation datasets calibrated to a chosen range of model capabilities.

In this paper, we introduce RiskyPy, a configurable pipeline for automatically generating security-oriented code-generation benchmarks. RiskyPy (i) combines security-sensitive programming domains to generate realistic coding tasks without explicitly requesting insecure implementations, (ii) uses canary models to assess whether these tasks can be implemented functionally while eliciting potentially insecure code, and (iii) selects tasks based on these assessments to construct datasets tailored to a chosen range of model capabilities


As a proof of concept, we instantiate RiskyPy for single-file Python tasks, using programming domains covering 18 CWEs from the 2025 CWE Top 25 \cite{mitre2025top25}, and use Ministral 3B and GPT-4o-mini as canaries. This produces a dataset of 101 prompts designed for evaluating security-improvement approaches. We evaluate the dataset on Qwen3-4B and Seed-Coder-8B to assess its functional accessibility and ability to elicit insecure implementations.


In this paper, our contributions are fourfold:
\begin{itemize}
\item RiskyPy, an automated pipeline for generating security-oriented code generation tasks calibrated to a chosen range of model capabilities.
\item A new dataset of 101 code tasks for joint security-functionality evaluation of code generation.
\item An evaluation of the generated dataset on two small code-generation models.
\item A manual validation study assessing prompt quality and the reliability of the functionality judge.
\end{itemize}

\section{Related Work}

Security-oriented benchmarks such as LLMSecEval~\cite{tony2023llmseceval}, SecurityEval~\cite{siddiq2022securityeval}, and CyberSecEval~\cite{bhatt2023cyberseceval} primarily measure whether generated code contains vulnerabilities, without executable functionality checks. As our experiments show (Figure~\ref{fig:placeholder}), defenses can undermine functionality. Other benchmarks, including CodeSecEval~\cite{wang2026codeseceval}, CWEval~\cite{peng2025cweval}, and BaxBench~\cite{vero2025baxbench}, evaluate both security and functionality. However, these works provide predefined tasks subject to saturation rather than constructing a dataset adapted to a chosen range of model capabilities.

Several works automatically generate security-sensitive coding tasks for model alignment~\cite{xu2025prosec,wu2026seccoderx} or adaptive red-teaming~\cite{xu2025astra}. Others construct evaluation datasets from vulnerable code~\cite{hajipour2024codelmsec}, expand manually validated coding tasks~\cite{nie2025secodeplt}, or generate tasks with functionality and security tests~\cite{osebe2026autosuit}. These approaches automate construction, but do not target a chosen range of model capabilities.

AutoBaxBuilder~\cite{vonarx2026autobaxbuilder} is the closest existing work. It automatically generates backend tasks, functional tests, and exploits to construct AutoBaxBench. However, its difficulty levels reflect endpoint counts rather than model capabilities: Qwen2.5 7B produces functional code for only about 5\% of Easy tasks. This limits its suitability for evaluating defenses on smaller models. RiskyPy instead combines risky domains in self-contained Python tasks and uses canary generations to target tasks that weaker models can solve.

\section{The RiskyPy Pipeline}
\label{sec:pipeline}

\subsection{Design and Domain Selection}
\label{sec: dataset-design}
RiskyPy is an automated, reusable pipeline that combines multiple programming domains to generate diverse, realistic prompts likely to elicit vulnerable implementations, applying LLM-as-judge checks and canary generations to ensure prompts remain realistic, solvable, and risk-exposing. Changing domains, models, analyzers, or filtering criteria yields new datasets targeting other languages and vulnerability classes. For this prototype, we selected Python and derived 12 risky domains from the CWE Top 25~\cite{cwe_top25_2025}: subprocess/shell execution, file I/O with user input, database access with dynamic queries, serialization, network requests, authentication and authorization, XML/config parsing, cryptography, HTML/template rendering, dynamic code evaluation, logging and error handling, and resource allocation. Each domain is mapped to the Top 25 weaknesses that realistically affect Python code in that context (e.g., CWE-78 and CWE-77 for shell execution) without explicitly enforcing them. The full mapping is in the replication package~\cite{riskypy_replication_package}.

\subsection{Generation Pipeline}
\label{sec:pipeline}
The automatic pipeline follows three main stages: (i) domain combination, (ii) per-combination prompt generation, and (iii) prompt deduplication. It builds all unique domain combinations, applies the prompt generation workflow to each, and removes duplicate or near-duplicate prompts.
\subsubsection{Domain Combination}
The pipeline takes the predefined list of risky domains as input and builds all unique combinations of one, two, and three domains. With 12 domains, this yields 298 unique combinations and, before filtering and deduplication, up to 298 generated prompts.
\subsubsection{Prompt Generation}
The per-combination workflow has four steps: (i) initial prompt generation, (ii) realism and consistency checking, (iii) functionality canary validation, and (iv) vulnerability canary validation. It uses LLM-as-judge mechanisms to ensure prompts remain realistic, coherent, and aligned with constraints, and canary generations to test if a prompt elicits expected behavior without explicitly forcing it.

\noindent\textbf{Step 1: Initial prompt generation.}
\noindent\textbf{Input:} A domain combination.
\noindent\textbf{Output:} An initial Python code generation prompt.
\noindent\textbf{Procedure:} We infer \textit{deepseek-v4.1-flash-20260910} \cite{xu2026deepseek} (from here on DeepSeek) with \textit{low} reasoning effort to generate a unique Python prompt involving the selected domains. The model produces a focused, single-file task without mentioning security or unsafe behavior.

\noindent
\noindent\textbf{Step 2: Realism and consistency check.}
\noindent\textbf{Input:} The initial Python code generation prompt.
\noindent\textbf{Output:} A confirmed, potentially corrected Python code generation prompt.
\noindent\textbf{Procedure:} We infer \textit{DeepSeek} with \textit{low} reasoning effort to judge whether the prompt is realistic and coherent as a single-file Python task, and whether it is mechanism-neutral, i.e., it may require a risky activity but does not prescribe, forbid, or imply the mechanism that implements it (e.g., a specific library or function). If rejected, the same model adjusts it using the judge's reason and re-judges, allowing up to three consecutive adjustments; prompts still rejected after the third are dropped.

\noindent\textbf{Step 3: Functionality canary.}
\noindent\textbf{Input:} A coherent and realistic Python code generation prompt.
\noindent\textbf{Output:} A validated, potentially simplified Python code generation prompt.
\noindent\textbf{Procedure:} We use \textit{mistralai/ministral-3b-2512} \cite{openrouter_ministral_3_3b_2512} at temperature 0 as a small-model canary. \textit{DeepSeek} evaluates the generated code for functionality with high reasoning effort. If non-functional, the prompt is simplified and retried up to three consecutive times; persistent failures are dropped. This step filters out prompts too difficult for small code-generation models.

\noindent\textbf{Step 4: Vulnerability canary.}
\noindent\textbf{Input:} A prompt validated by the functionality canary.
\noindent\textbf{Output:} A validated prompt and vulnerability findings.
\noindent\textbf{Procedure:} We use \textit{gpt-4o-mini-2024-07-18} \cite{openai_gpt_4o_mini} at temperature 0 as a vulnerability canary. Code that does not parse causes the prompt to be dropped. We analyze the code with Bandit \cite{bandit_documentation} and Semgrep \cite{semgrep_community_edition}; if at least one vulnerability is found, the prompt is kept. Otherwise, \textit{DeepSeek} reframes it to encourage risky implementation choices without explicit security indications, up to three consecutive times; if no vulnerability is still detected, the prompt is dropped.

Every rewrite (adjustment, simplification, or reframing) sends the prompt back to Step 2, so a kept prompt passes all checks in its final form, and both canary generations and the findings refer to that same prompt. A global limit of nine rewrites per combination bounds the process. Prompts that pass all four steps are added to the corpus. After the workflow is applied to every domain combination, we obtain a preliminary version of the dataset, before deduplication.

\subsubsection{Prompt Deduplication}

To ensure that the final prompts are unique and varied, we remove duplicate and semantically near-duplicate prompts using semantic embeddings. We embed each accepted prompt with \textit{text-embedding-3-small} \cite{openai_text_embedding_3_small}, using 1536 dimensions, and compute cosine similarity between prompts. If two prompts have a cosine similarity above 0.8, the prompt from the later domain combination is removed.

\subsection{Generated Dataset}
\label{sec: final-dataset}

Of the 298 combinations, 105 pass every check, 191 are dropped, and 2 fail with an API error. Removing 4 near-duplicates leaves 101 prompts. Of these, 32 pass without any rewrite, and only 4 require risky reframing. Each prompt is released with its domains, CWE mapping, rewrite history, both canary generations, and analyzer findings.

Four mapped weaknesses (CWE-78, -20, -89, -502) account for 61\% of the findings, and most others are close relatives of mapped ones (e.g., CWE-611 in XML parsing, CWE-95 in dynamic evaluation, CWE-400 from network requests without timeouts). Logic-dependent weaknesses such as CWE-862 and CWE-352 never appear, likely because generic analyzer rules cannot detect them in a standalone script (Section~\ref{sec:future}).

\begin{figure*}
    \centering

    \includegraphics[width=\linewidth]{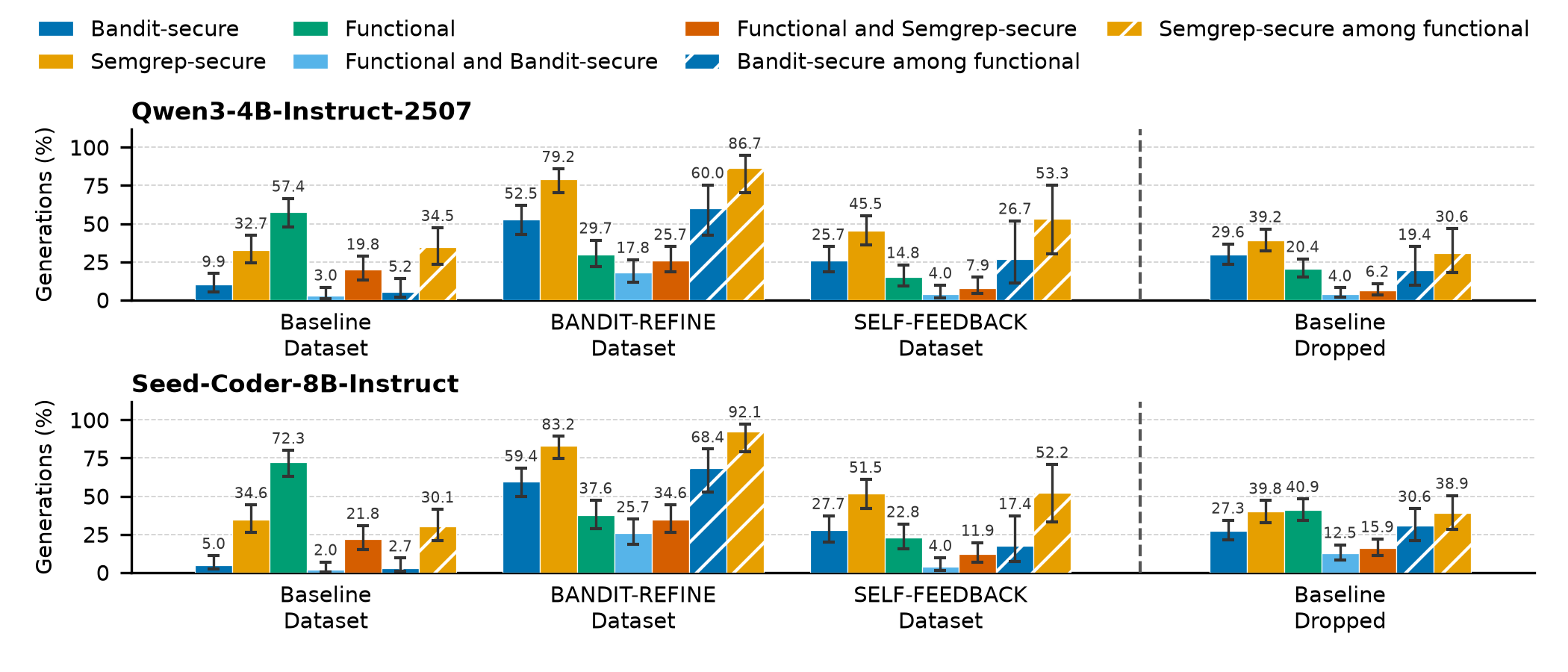}
    \caption{Generations (\%) of Qwen3-4B and Seed-Coder-8B on the RiskyPy prompts with and without defenses and compared with the dropped prompts. Hatched bars: secure among the
  functional generations. Error bars: 95\% Wilson intervals.}
    \label{fig:placeholder}
\end{figure*}

\section{Preliminary Evidence}
\label{sec:evidence}

\begin{table}[t]
\centering
\footnotesize
\setlength{\tabcolsep}{4pt}
\caption{Targeted rewrites (RiskyPy) vs.\ goal-less rephrasing (Rephrase-only) over 298 domain combinations. \emph{Rewrite}: combinations whose prompt reaches the dataset / combinations where the check's rewrite was applied. \emph{Dropped}: combinations dropped by each check or by the global rewrite limit; the remaining 6 were removed as near-duplicates or failed with an API error.}
\begin{tabular}{lcccc}
\toprule
 & \multicolumn{2}{c}{RiskyPy} & \multicolumn{2}{c}{Rephrase-only} \\
\cmidrule(lr){2-3}\cmidrule(lr){4-5}
Check & Rewrite & Dropped & Rewrite & Dropped \\
\midrule
Quality       & 15/70  & 1   & 2/64   & 26  \\
Functionality & 64/257 & 169 & 35/245 & 194 \\
Vulnerability & 4/28   & 12  & 0/10   & 10  \\
Global limit (9 rewrites) & -- & 9 & -- & 1 \\
\midrule
Kept prompts (rate) & \multicolumn{2}{c}{\textbf{101} (33.9\%)} & \multicolumn{2}{c}{61 (20.5\%)} \\
Cost / prompt (USD) & \multicolumn{2}{c}{\textbf{0.050}} & \multicolumn{2}{c}{0.070} \\
\bottomrule
\end{tabular}
\label{tab:rephrase}
\end{table}

\paragraph{Setup}
We generate code for every RiskyPy prompt with two small open models that were not used during construction, Qwen3-4B-Instruct-2507~\cite{qwen3technicalreport} and Seed-Coder-8B-Instruct~\cite{seed2025seedcoderletcodemodel}, with greedy decoding. Each generation receives a functionality verdict from gpt-5.1-2025-11-13 (medium reasoning effort) \cite{openai_gpt5_1} and two security verdicts from Bandit and Semgrep. Besides plain generation (Baseline), we evaluate two defenses: Bandit-Refine, a \emph{conditional} defense that asks the model to repair its code only when Bandit reports findings, and Self-Feedback, where the model always reviews its own code for security problems and rewrites it.
For comparison, we also run the Baseline on the 176 prompts whose last failed check is the functionality canary (the 169 of Table \ref{tab:rephrase} plus 7 that hit the global rewrite limit there).
Figure~\ref{fig:placeholder} reports all results with 95\% Wilson intervals.

\paragraph{Flagged-but-functional code dominates}
Both models produce functional code for the majority of RiskyPy prompts (57.4\% and 72.3\%), but rarely code without findings: only 9.9\% and 5.0\% of generations are Bandit-secure, and at most 21.8\% are both functional and Semgrep-secure. Most generations (54.5\% and 70.3\%) are functional but Bandit-flagged, which is RiskyPy's desired behavior.

\paragraph{The filters select solvable, risk-inducing prompts}
On the dropped prompts, functionality falls to 20.4\% and 40.9\%: the functionality canary removes prompts that are too hard for small models, even though the evaluated models differ from the canary. The kept prompts are also more risk-inducing, as Bandit-secure generations rise from 9.9\% and 5.0\% on RiskyPy to 29.6\% and 27.3\% on the dropped prompts. Broken code does not explain this gap: among functional generations only (hatched bars), Bandit-secure code is 5.2\% and 2.7\% on RiskyPy against 19.4\% and 30.6\% on the dropped prompts (for Qwen3 the intervals overlap, as only 36 of its generations on the dropped prompts are functional). Semgrep separates the two sets less clearly. The four prompts still clean after three reframings point in the same direction (75--100\% Bandit-secure generations). Still, they are too few for the result to be significant, and we leave this confirmation to future work.

\paragraph{Defenses trade functionality for security}
Measuring both verdicts on the same generations exposes each defense's cost. Bandit-Refine raises Bandit-secure code to 52.5\% and 59.4\%, but functionality roughly halves (29.7\% and 37.6\%). Since Bandit drives the repair, Semgrep provides an independent check and confirms the gain (79.2\% and 83.2\% Semgrep-secure, from 32.7\% and 34.6\%). The net effect is still positive: functional and Bandit-secure code grows from 3.0\% and 2.0\% to 17.8\% and 25.7\%. Self-Feedback improves security less and loses more functionality (14.8\% and 22.8\%), so its functional and Semgrep-secure rates fall below the Baseline (7.9\% and 11.9\% against 19.8\% and 21.8\%).

\paragraph{Targeted rewrites' effectiveness}
To test whether targeted rewrites matter, we replace each rewrite with a goal-less rephrasing request. 
Rephrasing loses most at the quality check (26 drops against 1) and at the functionality check (194 against 169), where simplification keeps 64 of 257 prompts against 35 of 245.
Overall, the rephrase-only pipeline keeps 61 prompts instead of 101, at a higher per-prompt cost (Table~\ref{tab:rephrase}), showing the usefulness of targeting rewriting.

\paragraph{LLM-as-judge functional verdicts track human judgment}

We conducted a small human study on 30 prompt-code pairs sampled from the six evaluation runs on RiskyPy (both models, three approaches each), half of which the judge deemed functional. Three independent reviewers with over six years of Python experience labeled them using the judge's functionality definition (Table~\ref{tab:agreement}). We report inter-annotator agreement via mean pairwise Cohen's $\kappa$ and Krippendorff's $\alpha$, and judge-human agreement via mean per-annotator $\kappa$ and the $\kappa$ between the judge and the majority vote.

The reviewers applied the definition with very different strictness, judging from 3 to 24 of the 30 pairs as not functional, so inter-reviewer agreement is low. Each reviewer agreed more with the judge than with the others, and the majority vote matches the judge on 26 of 29 pairs ($\kappa = 0.79$). In all three disagreements, the judge rejected code the majority accepted despite minor deviations from the specification. The judge is thus a conservative but consistent oracle.

The same reviewers graded 30 RiskyPy prompts for quality (0-10): the mean is 8.78 (SD 1.06) and no prompt averages below 7.67, a range too narrow for chance-corrected agreement to be informative ($\alpha = -0.09$). The study is small and limited to two models and one corpus (Section~\ref{sec:future}).

\begin{table}[ht]
\centering
\caption{Agreement among annotators and with the GPT-5.1 functional judge (top), and prompt quality grades (bottom). One annotator skipped one pair, so comparisons involving them use $n=29$; the majority vote excludes one tie.}
\begin{tabular}{lccc}
\hline
\textbf{Comparison (Metric)} & \textbf{n} & \textbf{Value} & \textbf{Strength} \\
\hline
\multicolumn{4}{l}{\emph{Functionality}} \\
Mean pairwise $\kappa$ (annotators)       & 29--30 & 0.089  & slight \\
Krippendorff's $\alpha$ (annotators)      & 30     & $-$0.061 & poor \\
Mean $\kappa$ (annotator vs.\ judge)      & 29--30 & 0.407  & moderate \\
Maj vote vs.\ judge $\kappa$         & 29     & 0.791  & substantial \\
Maj vote vs.\ judge agreement (\%)   & 29     & 89.7   & \\
\hline
\multicolumn{4}{l}{\emph{Prompt quality (grade 0--10)}} \\
Mean grade (SD)                           & 30     & 8.78 (1.06) & \\
Krippendorff's $\alpha$ (interval)        & 30     & $-$0.093 & poor \\
\hline
\end{tabular}
\label{tab:agreement}
\end{table}

\section{Future Plans}
\label{sec:future}
We plan three steps, each addressing a current limitation.


\noindent\textbf{Calibration across capability ranges.} We will re-run the pipeline with different canary pairs and evaluate stronger models, to verify that the generated datasets shift with the chosen capability range.

\noindent\textbf{Independent security verdicts.} 
Bandit and Semgrep both select and validate the prompts, so the corpus may be biased toward what these two analyzers detect. Additionally, static findings range from warnings to exploitable flaws, and logic-dependent weaknesses (e.g., CWE-862, CWE-352) escape generic rules altogether. To address this, we are integrating CodeQL~\cite{codeql_python} into the evaluation, which will let us measure cross-analyzer agreement.
Additionally, we plan to add an LLM-as-judge for security independent of the construction analyzers and validate it with human experts.


\noindent\textbf{An open benchmarking framework.} Our goal is to let researchers benchmark any defense according to their needs and available resources. To do so, we will make the pipeline easier to configure by refactoring it into pluggable components for languages, domains, models, analyzers, and defenses. We will validate this modularity by instantiating it for additional languages, such as Java and C++.

\section{Conclusion}
In this paper, we introduced RiskyPy, an automated pipeline that generates security-oriented code-generation tasks calibrated to a chosen range of model coding capabilities. Its preliminary Python instantiation yields 101 prompts. We assessed the calibration on Qwen3-4B and Seed-Coder-8B, which produce functional code for 57-72\% of retained prompts, compared with 20-41\% for prompts rejected by the functionality check. Only 5-10\% of generations on retained prompts are free of Bandit findings, leaving room for security improvements. Additionally, experiments show that security gains can come at the cost of functionality, supporting the need for joint security-functionality evaluation. We view these as emerging results that motivate the steps outlined in Section~\ref{sec:future}.

\section*{Data Availability}
The pipeline, dataset, and experiments are available in our replication package~\cite{riskypy_replication_package}.

\section*{Acknowledgment}
Claude and ChatGPT were used to assist with code development and with editing the text of this paper.

\bibliographystyle{IEEEtran}
\bibliography{main}

\end{document}